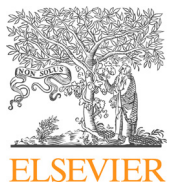

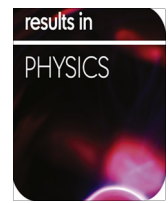

# Exploring the effects of photon correlations from thermal sources on bacterial photosynthesis

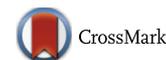

Pedro D. Manrique [a,*], Felipe Caycedo-Soler [b], Adriana De Mendoza [c], Ferney Rodríguez [c], Luis Quiroga [c], Neil F. Johnson [a]

[a] Physics Department, University of Miami, Coral Gables, FL 33126, USA
[b] Department of Physics, Universitat Ulm, Germany
[c] Departamento de Física, Universidad de Los Andes, Bogotá, Colombia

## ARTICLE INFO

## ABSTRACT

Thermal light sources can produce photons with strong spatial correlations. We study the role that these correlations might potentially play in bacterial photosynthesis. Our findings show a relationship between the transversal distance between consecutive absorptions and the efficiency of the photosynthetic process. Furthermore, membranes where the clustering of core complexes (so-called RC-LH1) is high, display a range where the organism profits maximally from the spatial correlation of the incoming light. By contrast, no maximum is found for membranes with low core-core clustering. We employ a detailed membrane model with state-of-the-art empirical inputs. Our results suggest that the organization of the membrane's antenna complexes may be well-suited to the spatial correlations present in a natural light source. Future experiments will be needed to test this prediction.

© 2016 The Authors. Published by Elsevier B.V. This is an open access article under the CC BY-NC-ND license (http://creativecommons.org/licenses/by-nc-nd/4.0/).

## Introduction

The harvesting of light by photosynthetic organisms constitutes the main source of energy that sustains life on Earth [1–8]. Both the intensity and statistics of light arriving to distant parts of the Universe may be quite different from our own Sun, particularly in terms of the inherent correlations within the light. It has been noted that thermal radiation from a distant source carries long-range spatial correlations in intensity [9,10]. In addition, the area of coherence associated with sunlight is estimated to be $\Delta A \approx 3.67 \times 10^{-3} \text{ mm}^2$ [11]. Therefore, one can reasonably postulate that photosynthetic organisms contained within this range might be expected to experience the effects of spatial coherence of natural sunlight. This raises the question of how bacteria might cope with, or adapt to such correlations in the incident light. It is this issue that we start to explore in this paper. While our results have as yet no firm experimental support, we hope that they stimulate future explorations along this line of research.

Among the wide variety and complexity of photosynthetic organisms, bacterial photosynthesis constitutes an ideal focus for research. It is arguably the oldest form of photosynthetic life [12–14], while its structure is simpler than higher organisms like plants and algae [2–4]. Advances in the study of bacterial photosynthesis through Atomic Force Microscopy (AFM) imaging have helped clarify the structure and organization of the photosynthetic apparatus [6,15]. The composition of harvesting structures has been observed to vary depending on the light intensity conditions in several purple bacteria species [16–18]. In addition, high resolution AFM images have discovered variations in core complex architectures across species. For example, it was discovered that species like Rsp. Palustris express a monomeric core complex while for Rb. blasticus and Rb. sphaeroides it is dimeric [19]. Although a single transmembrane helix protein PufX may be responsible for inducing dimerization of the core complex [20], it is still unclear what benefits this development brings from the point of view of photon absorption [21,22]. Furthermore, species like Rsp. photometricum naturally express clusters of core complex. The samples analyzed in Ref. [23] show that 73% of core complexes make a maximal number of three core-core contacts. Potential benefits have been studied from the point of view of photo-excitation migration [24] but from the point of view of photon absorption, they still lack quantification.

In this paper, we analyze the potential role of spatial correlations in the photon absorption on the metabolism of the photosynthetic organism. Such correlations will naturally be present in most light sources but have been assumed to be too small to make an impact. We examine the photosynthetic efficiency, which is the amount of solar energy absorbed by the antenna complexes that is then transformed into chemical energy in the reaction center (RC). Our theoretical model has already been shown to capture the chromatic adaptation of Ref. [17] in terms of the dynamic interplay between the excitation kinetics and the RC cycling [25,26]. Specifically, we show that spatial correlations in the statistics of photon arrival and hence absorption, could yield an enhancement of the efficiency of the photosynthetic process. Moreover, our results show that the core-core clustering plays a fundamental role on the degree of this enhancement. In particular, membranes with high core-core clustering display higher sensibility to the spatial correlations when compared with disordered membranes.

## Photosynthetic membrane model

The photosynthetic purple bacteria use light harvesting complexes (LHC) that are spatially distributed on the cytoplasmic membrane [3,7,27,28] to capture incoming photons within a specific range of the spectra. The complex referred to as Light Harvesting 1 (LH1) generally absorbs maximally at 875 nm while the complex Light Harvesting 2 absorbs maximally at 800 nm and 850 nm [3,29,30]. The absorbed photo-excitation is then transferred to the photosynthetic reaction center (RC) or dissipated at specific transfer and dissipation rates, respectively. A set of pigments in the RC undergo a reduction after electronic excitation. Two of these electrons are necessary to form a molecule of Quinol ($Q_BH_2$), therefore two photo-excitations are required. Once a RC receives the second photo-excitation it takes a few milliseconds to produce Quinol, undock the new molecule and be substituted by a new Quinone ($Q_B$). Absorption rates have been calculated for LH1 and LH2 and normalized to a light intensity of 1 W/m² resulting in $\gamma_1 = 1$ s⁻¹ for LH1 and $\gamma_2 = 0.55$ s⁻¹ for LH2 [25,31]. In this way, a membrane with $N_1$ LH1 complexes and $N_2$ LH2 complexes has an absorption rate $\gamma_A = I(\gamma_1 N_1 + \gamma_2 N_2)$ for a given light intensity $I$. The transfer rates measured from pump–probe experiments are in good agreement with generalized Föster calculations [32], assuming intra-complex delocalization. LH2→LH2 transfer rate has been calculated as $t_{22} = 10$ ps [32], while LH2→ LH1 transfer has been measured for *R. Sphaeroides* as $t_{21} = 3.3$ ps [33]. Back-transfer LH1→ LH2 is approximately $t_{12} = 15.5$ps while the LH1→ LH1 mean transfer time $t_{11}$ has been calculated using a generalized Förster interaction [5] as 20 ps. Second and third lowest exciton lying states cause LH1→ RC transfer [34], consistent with a transfer time of 35 – 37 ps found experimentally at 77 K [35,36]. As proposed in Ref. [27], increased spectral overlap at room temperature improves the transfer time to $t_{1,RC} = 25$ ps. The back-transfer from an RC's fully populated lowest exciton state to higher-lying LH1 states occurs in a calculated time of $t_{RC,1} = 8.1$ ps [34], which is close to the experimentally measured 7–9 ps estimated from decay kinetics after RC excitation [37]. The subsequent passage through the RC complex depends on whether the RC is neutral (i.e. the RC is in an open state), and typically occurs within $t_+ = 3$ ps. The dissipation rate $\gamma_D$ resulting from fluorescence and internal conversion mechanisms, has been estimated to be $\gamma_D = 1$ ns⁻¹ [5]. Our model adopts a stochastic approach to the classical rate equations for a large number of LHC ($\approx$400). It accounts for photon absorption, photo-excitation transfer, dissipation and RC cycling for a given architecture and light statistics.

Figure 1 illustrates schematically our theoretical model. At each time step ($\delta t \approx 0.025$ ps) incoming photons are absorbed at a combined rate $\gamma_A$ by complexes LH1 and LH2. Our semiclassical model places consecutive absorptions within a correlation radius $r$ in order to simulate the spatial bunching of thermal light. When $r$ is large enough to cover the whole membrane, we recover the limit of random absorption (no correlation). This simple mechanism allows us to model the grouping effect present in pairs of photons, where the bunching degree is related to the inverse of the parameter $r$. Consequently, the absorbed photo-excitation diffuses throughout the membrane in search for an open RC according to transfer rates for a given membrane architecture. Once an open RC has received two photo-excitations, it is set closed and no other photo-excitation is allowed to enter. After a time $\tau_{RC}$ has elapsed from the moment in which the second photo-excitation has entered, the RC is set open and the cycle starts from the beginning. This open/close mechanism accounts for the time where two electrons produce $Q_BH_2$ before it undocks and a new $Q_B$ substitutes it. This cycling process lasts a few milliseconds and successfully explains structural preferences in adaptation of purple bacteria [16–18,25,26].

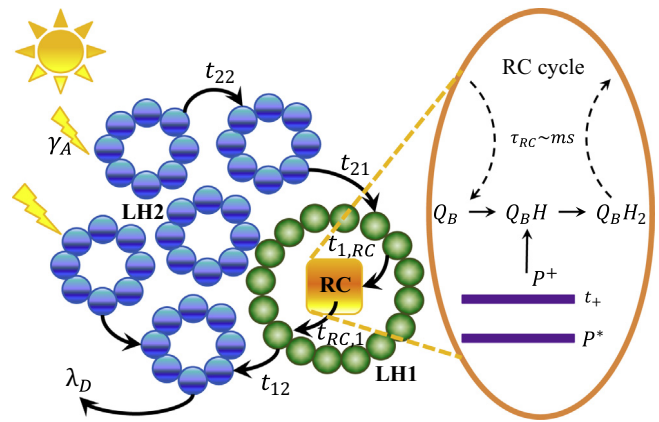

**Fig. 1.** Schematic of our theoretical model for photon absorption and excitation transfer. Radiation from a thermal source (as approximated by our Sun, for example) is absorbed by LHC complexes 1 (blue rings) or 2 (green rings) at a combined rate $\gamma_A$. The resulting excitation is transferred from complex i to complex j at a specific rate $t_{ij}$ or it is dissipated at a rate $\lambda_D$. At the reaction center, the double excitation of the special pair $P$ yields to the formation of quinol ($Q_BH_2$) molecule in a cycle that lasts a few milliseconds. (For interpretation of the references to color in this figure legend, the reader is referred to the web version of this article.)

## Results and discussion

The membrane's performance is quantified by its efficiency $\eta$, defined as the ratio of photo-excitations that produce charge separation in the RCs, to the total number of absorbed photons. Similarly we define the relative efficiency $\eta/\eta_{RA}$, to compare the efficiency $\eta$ of correlated photons and the efficiency $\eta_{RA}$ for the scenario where spatial correlations are not accounted for. We study membranes which have the same number of LH1 and LH2 complexes ($N = 400$), but differ in the specific architecture/configuration of complexes; in particular, the RC-LH1 core-core clustering. Hence the contrasting situations of high core-core clustering (HCC, Fig. 2(a)) and sparse low core-core clustering (LCC, Fig. 2 (b)), differ in terms of the mean number of RC-LH1 which encircle any RC-LH1 core-complex. Fig. 2(a) represents well the maximal core-core clustering that is observed in *Rsp. photometricum* for low-light intensity growth [23] with a linear core arrangement similar to *Rb. sphaeroides* [38], while Fig. 2(b) resembles the vesicles grown under high-light intensity in *Rsp. photometricum* [23].

Fig. 2(c) shows the striking result that HCC membranes benefit from the spatial correlations present in the incident sunlight. Our model captures a peak enhancement in the photosynthetic efficiency for membranes with HCC clustering (blue triangles), which is however absent in the LCC (red circles). Our simulations reveal that HCC presents a reduction in the number of complexes visited before charge separation takes place, when consecutive absorptions occur within radii compatible with the core clusters size. This can be understood as a benefit of placing clustered cores in the presence of the incident thermal light which is spatially bunched. Therefore it is desirable to envision clustered charge separation units for efficient conversion in artificial devices subject to sunlight excitation. Our simulation shows the output for a RC closed time of 12.5 ms in agreement with the average time for quinone release [39,40], however values of closed times up to 50 ms have proven to preserve the observed effect. By contrast, for negative spatial correlation in the incident light (i.e. opposite of natural sunlight), the peak disappears as shown in Fig. 2(d).

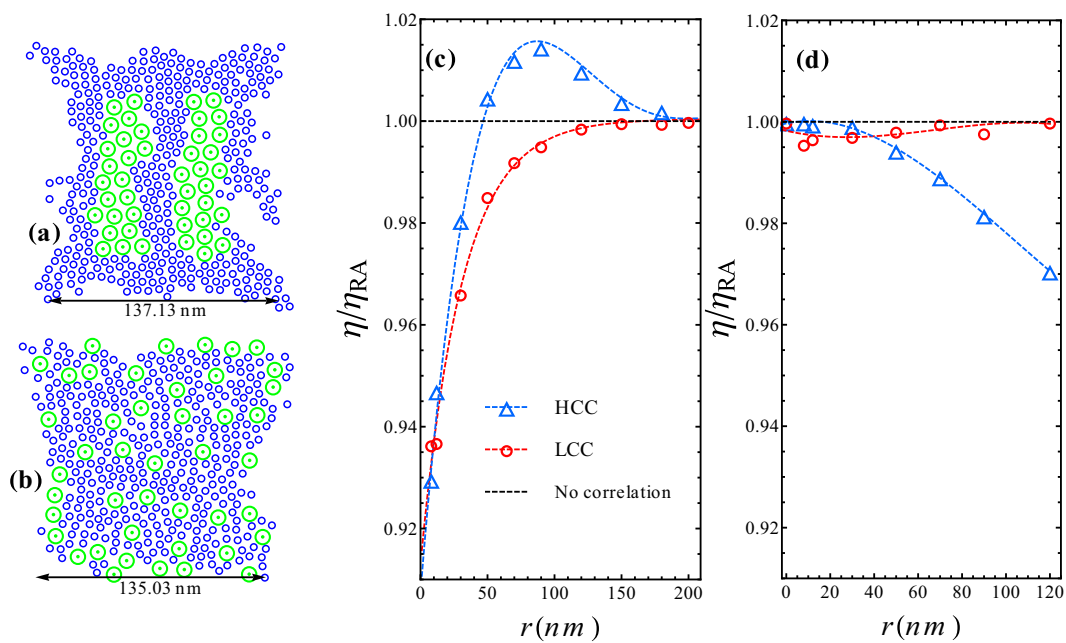

**Fig. 2.** Effects of light correlations on photosynthetic relative efficiency according to their core-core clustering. Architectures of the antenna complexes (a) HCC and (b) LCC. They are obtained as local minima in a large-scale Monte Carlo energy minimization, and are hence realistic as (locally stable) minimum energy structures. Blue rings represent LH2 antennae and green rings are core RC-LH1 complexes. (c) Photosynthetic relative efficiency as a function of the correlation parameter $r$ for architectures HCC (triangles) and LCC (circles). Dashed black line illustrates the result without light correlation. Symbols represent the result from our model while lines correspond to the fitting of our simulation points. (d) Analogous to (c) but for negative correlation. The RC closure time is $\tau_{RC} = 12.5$ ms. (For interpretation of the references to color in this figure legend, the reader is referred to the web version of this article.)

While other energetic and metabolic factors (e.g. complexes affinity and charge carrier diffusion) may ultimately dictate a given membrane's architecture in a given environment, our results also speak to open questions about the purpose of membrane organization [38,41], showing that it allows the system to harvest not only the light's energy, but also to profit from its spatial correlations. The increment in efficiency shown in Fig.2(c) should not be regarded as negligible: It has been demonstrated that for these organisms, large structural variations occur as a result of seemingly modest metabolic benefits. For example, *Rsp. acidophilia* develops an expensive structural adaptation under low intensity illumination that gradually replaces LH2 by LH3 complexes which, due to a red shifted maximum, improves the transfer to the LH1 which in turn improves the membrane's efficiency by 3–4% [30].

## Conclusions

We have used a simple yet empirically grounded model to study the potential effects of spatial correlations on bacterial photosynthesis, for membrane architectures with contrasting core-core clustering. We have shown that clustering of charge separation units might in principle be used to exploit the spatial correlation of thermal light by reducing the mean free path required for absorbed excitations to reach a charge separation unit. These results suggest that the photosynthetic organism may be able to profit from the incident light's implicit spatial bunching in order to satisfy its metabolic needs.

## Acknowledgement

We thank for partial financial support: COLCIENCIAS (Colombia) through the grant for doctoral studies in Colombia (call 528); Banco de la Republica de Colombia through the project 3646 and Faculty of Science; Universidad de Los Andes, Colombia (call 2014-2). F.C.S. acknowledges support from the EU project PAPETS, the ERC Synergy grant BioQ and the DFG via the SFB TR/21. N.F.J. is funded by National Science Foundation (NSF) Grant CNS1522693 and Air Force (AFOSR) Grant FA9550-16-1-0247.